\documentclass[11pt,a4paper,reqno]{amsart}
\usepackage[dvipsnames]{xcolor}
\usepackage{siunitx}
\usepackage{graphicx} 
\usepackage[headings]{fullpage}
\usepackage{amsaddr}
\usepackage{dsfont}
\usepackage[singlelinecheck=false]{caption}
\usepackage{url}
\usepackage{float}
\usepackage{multirow}
\usepackage{mathtools}
\usepackage{upgreek}
\usepackage{subcaption}
\usepackage{amssymb}
\usepackage{amsmath}
\usepackage{tikz}
\usepackage{hyperref}
\usepackage{siunitx}
\usepackage{matlab-prettifier}
\usepackage[noadjust]{cite} 
\usepackage[version=4]{mhchem}
\usepackage[toc,page]{appendix}

\usetikzlibrary{shapes, arrows, positioning,decorations.markings, arrows.meta}
\usetikzlibrary{arrows, arrows.meta, calc, positioning, quotes, shapes}
\makeatletter
\renewcommand{\section}{\@startsection
  {section}{3}{0mm}%
  {-\baselineskip}{0.5\baselineskip}%
  {\centering\normalfont\Large\bfseries}}%
\makeatother

\makeatletter
\renewcommand{\subsection}{\@startsection
  {subsection}{3}{0mm}%
  {-\baselineskip}{0.5\baselineskip}%
  {\normalfont\bfseries}}%
\makeatother

\makeatletter
\renewcommand{\subsubsection}{\@startsection
  {subsubsection}{3}{0mm}%
  {-\baselineskip}{-0.5\baselineskip}%
  {\normalfont\itshape\bfseries}}%
\makeatother

\makeatletter
\renewcommand\paragraph{\@startsection{paragraph}{4}{\z@}%
                                    {1ex \@plus1ex \@minus.2ex}%
                                    {-1em}%
                                    {\normalfont\normalsize\itshape}}
\makeatother

\newcommand{\code}[1]{\texttt{#1}}

\begin{document}

\title{Tracking the Emergence and Persistence of Abnormal Grain Growth in an Aluminum Alloy using Diffraction-Contrast Tomography}

\author[]{Helmuth-André Schulz-Harder$^1$, Karolína Gutbrod$^1$, Madlen Atzen$^1$, Markus Ziehmer$^1$, Thomas Wilhelm$^2$, Orkun Furat$^{2,3}$, Yubin Zhang$^4$, Jette Oddershede$^5$, Haixing Fang$^6$, Pierre-Olivier Autran$^6$, Wolfgang Ludwig$^{6,7}$, Marc Bernacki$^8$, Volker Schmidt$^2$, Carl E. Krill III$^1$, Jules M. Dake$^1$}
\address{$^1$Institute of Functional Nanosystems, Ulm University, 89081~Ulm, Germany}
\address{$^2$Institute of Stochastics, Ulm University, 89069~Ulm, Germany}
\address{$^3$SDU Applied AI and Data Science Unit, University of Southern Denmark,\\ 5230~Odense, Denmark}
\address{$^4$Department of Civil and Mechanical Engineering, Technical University of Denmark, DK-2800 Kgs. Lyngby, Denmark}
\address{$^5$Xnovo Technology ApS, Galoche Allé 15, 1st floor, 4600, Køge, Denmark}
\address{$^6$European Synchrotron Radiation Facility, 71 Avenue des Martyrs, 3800~Grenoble, France}
\address{$^7$Université de Lyon, INSA Lyon, CNRS MATEIS, 69621~Villeurbanne, France}
\address{$^8$Mines Paris, PSL University, Centre for Material Forming (CEMEF), UMR CNRS, 06904~Sophia Antipolis, France}

\begin{abstract}
Abnormal grain growth is inherently three-dimensional, making 3D tracking of grain evolution essential for understanding its underlying mechanisms. This study combines synchrotron-based diffraction-contrast tomography and phase-contrast tomography to characterize the three-dimensional grain structure and quantify the size and spatial distribution of second-phase particles in the aluminum alloy AA5252. Using an interrupted annealing protocol, we tracked the persistent growth of abnormal grains and the emergence of abnormal growth in a grain initially belonging to the matrix. Although second-phase particles effectively suppressed grain growth in the matrix, variations in local particle density showed no apparent correlation with either the emergence or subsequent rapid growth of abnormal grains. Instead, crystallographic misorientation and, more notably, neighboring grain size were identified as stronger predictors of abnormal growth rate. Matrix grains were consumed more rapidly when they were relatively small and shared high-angle grain boundaries with the growing abnormal grain. The same trends were observed for the grain that underwent a transition from normal to abnormal growth. Remarkably, even before the onset of abnormal growth, this grain was surrounded by smaller matrix grains with exclusively high-angle grain boundaries, while the local particle density was slightly higher than average. These findings suggest that the same microstructural characteristics facilitate both the emergence and persistence of abnormal grain growth, although the mechanism triggering the transition to rapid growth remains unresolved.
\end{abstract}

\maketitle

\section{Introduction} \label{intro}
Heating a polycrystalline material almost inevitably leads to coarsening of its underlying microstructure. In single-phase materials, the mechanism of coarsening is referred to as grain growth, and when this process proceeds in a more-or-less uniform manner, it is called normal grain growth. Under certain experimental conditions, normal grain growth can be suppressed such that only a subset of grains in the polycrystal is able to grow. These grains rapidly consume their slower-growing or completely stagnant neighbors (the grains making up the ``matrix''), and the resulting growth mode is known as abnormal grain growth (AGG).

AGG is associated with a transition from a unimodal to a bimodal grain size distribution (GSD) \cite{dunn_secondary_1966, humphreys_recrystallization_2017}. Such bimodality can significantly alter the properties of crystalline materials, i.e., most metals and ceramics. Depending on the application, AGG may be desired, but usually it is not. For example, in grain-oriented silicon steels \cite{Park_Fe_Si_2018}, AGG helps to develop a magnetically preferred orientation, which reduces core losses in electrical transformers. On the other hand, AGG is generally undesirable in steels, as it leads to a heterogeneous microstructure characterized by a small number of excessively large grains \cite{gladman_abnormal_1992}. Such microstructural heterogeneity can compromise mechanical performance, as demonstrated in friction-stir-welded low-carbon steel, where AGG has been associated with reduced hardness and tensile strength \cite{Sun2017}. Similar detrimental effects have been reported in other engineering alloys. In the nickel-based superalloy Inconel 718, which is widely used in aerospace applications due to its high strength and excellent fatigue resistance, AGG has been shown to significantly reduce fatigue life \cite{AGG_inconel718}. Beyond its effects on bulk mechanical properties, AGG can also impair surface quality. In 6xxx-series aluminum alloys used for lightweight automotive body panels, the formation of a few abnormally large grains within a finer-grained matrix can promote surface defects such as roping \cite{ouhiba_6016_aluminium}.

Phenomenologically, we know that, in order for AGG to occur, a certain number of grains must first obtain and then somehow retain a growth advantage over crystallites in the surrounding matrix. Any proposed mechanism for AGG, therefore, must be able to explain an abnormal grain's \textit{emergence}---the moment it begins outgrowing its neighbors---as well as its \textit{persistence} in growing faster. (Here, we prefer to use emergence rather than nucleation to avoid confusion with phenomena such as recrystallization or phase transformations, for which nucleation is associated with an energy barrier.)

Of the many mechanisms for AGG mentioned in the literature \cite{krill_iii_extreme_2023}, the two most commonly cited are related to the presence of texture \cite{rollet_texture_AGG_2005, Kim_Al_texture_AGG_2007, humphreys_recrystallization_2017} and pinning particles \cite{dennis_abnormal_2009, holm_particle-assisted_2015}. In textured materials, AGG can occur when a few grains deviate from the main component of crystallographic orientation in the specimen \cite{humphreys_recrystallization_2017}. Because high-angle grain boundaries (HAGBs) typically have higher energies \cite{read_dislocation_1950} and mobilities \cite{humphreys_unified_1997_1} than low-angle grain boundaries (LAGBs), any above-average-sized misaligned grain that forms HAGBs with matrix grains belonging to the main component of crystallographic orientation can outgrow the surrounding matrix, whose grains are predominantly separated by LAGBs owing to the crystallographic texture. Similarly, AGG can also be observed in materials containing small, second-phase particles that retard or even completely prevent the migration of grain boundaries through Zener pinning \cite{smith_grains_1948, Gladman1966, manohar_five_1998}. According to Humphreys' model for AGG \cite{humphreys_unified_1997_2}, the largest grains in a pinned microstructure may continue to grow as long as they exceed their neighbors in size by a sufficiently large ratio, whereby the exact value of this threshold depends on the size and volume fraction of the pinning particles. In this scenario, persistence follows from the driving force for grain growth being able to overcome the particle pinning pressure whenever a sufficiently large grain comes into contact with a smaller matrix grain \cite{humphreys_unified_1997_2}. The growth advantage is maintained in this mechanism by the fact that, as a large grain grows, its size ratio relative to the matrix steadily increases, provided the matrix grains remain (nearly) pinned.

While AGG has long been studied in two-dimensional (2D) images of microstructures, only a handful of time-resolved three-dimensional (3D) measurements have appeared to date in the literature \cite{lu_dynamics_2020, zhu_3d_2023, wang_4d_2025, DeLellis_AGGalumina_2026, Ånes_2026}. This scarcity persists despite the growing availability of nondestructive 3D imaging techniques \cite{poulsen_three-dimensional_2001, ludwig_x-ray_2008, lienert_high-energy_2011, liu_three-dimensional_2011, sorensen_multigrain_2012, king_first_2013, labDCTHolzner_2016}. Because AGG is an inherently 3D phenomenon, such measurements are essential to capture the emergence of abnormal grains and to gain deeper insight into the mechanisms driving their persistent growth during prolonged annealing.

The proper characterization of these dynamics, however, is complicated by certain aspects of AGG itself. Even after conditions promoting AGG have been established, abnormal grains may fail to form within a given region under investigation, or their growth may proceed too quickly between annealing steps to reveal the trajectory taken by the abnormal grain through the matrix, obscuring the mechanisms involved. Moreover, technical difficulties in reconstructing polycrystalline volumes with bimodal grain size distributions often lead to an inaccurate determination of grain boundary (GB) locations precisely where the latter are most relevant to AGG:\ namely, at the interfaces between abnormal and matrix grains. The aforementioned challenges are even more pronounced in materials exhibiting extreme forms of AGG, during which abnormal grains grow to sizes at least an order of magnitude larger than the mean size of the matrix grains \cite{krill_iii_extreme_2023}. Conversely, it is precisely these materials that enable us to focus on the clearest cases of AGG, where one might expect the underlying mechanism(s) to be easiest to detect.

In light of these considerations, we employed synchrotron-based diffraction-contrast tomography (DCT) at beamline ID11 of the European Synchrotron Radiation Facility (ESRF) to map a representative volume of the commercial aluminum alloy 5252 (AA5252) following sequential discrete (15-minute) annealing steps. This material is known to exhibit extreme AGG in a certain range of annealing temperatures, thus raising the likelihood of detecting incipient AGG and capturing the subsequent persistent growth of individual abnormal grains. DCT is capable of delivering high-resolution mappings of polycrystalline microstructures \cite{ludwig_three-dimensional_2009, zhang_three-dimensional_2018}, even for fine-grained materials, which enables individual GBs to be located and tracked during annealing-induced migration. Furthermore, the high spatial coherence of the x-ray beam at synchrotron sources makes phase-contrast tomography (PCT) \cite{zoofan_phase-contrast_2006} feasible, by which second-phase particles---known to be present in AA5252---can be identified, even when their absorption contrast is too low for conventional absorption-based computed tomography (CT). By fusing DCT and PCT datasets, we can examine how grain neighborhood relationships, GB character (in particular, crystallographic misorientation) and second-phase particles influence the emergence and persistence of AGG in AA5252.

\section{Material and methods}\label{methods}
Selecting a material that exhibits AGG and can be annealed in ambient air simplifies microstructure characterization. We therefore chose the commercial aluminum alloy AA5252, which, based on preliminary measurements, manifests AGG at annealing temperatures around $\SI{500}{\celsius}$. This alloy is closely related to AA5052, which has a comparable chemical composition within the Al–Mg (5xxx) series and has been studied extensively with regard to AGG \cite{suk_development_2006, park_ex_2012, park_new_2015, na_misorientation_2016}. The larger mean grain size of recrystallized AA5252, however, improves the reliability of 3D characterization by DCT and was an important factor in selecting this alloy. In both alloys, the alloying elements promote the formation of second-phase particles \cite{engler_impact_2018, Gao_defectsAA5252_2020}, providing an opportunity to investigate their role in particle-mediated AGG \cite{holm_particle-assisted_2015}.

\subsection{Sample preparation and data acquisition}
A cylindrical specimen with a diameter of $\SI{0.8}{\milli\meter}$ was prepared by wire electrical discharge machining of a cold-rolled AA5252 sheet (80\% thickness reduction), which was supplied by Constellium and rolled at the Centre for Material Forming (CEMEF). Magnesium is the primary alloying element in 5xxx series aluminum alloys, and our sample contained approximately 2.7\,wt\% Mg and 0.4\,wt\% Mn, together with smaller amounts of Fe, Si, Zn and Cu. While Zn and Cu remain largely in solid solution, Mn, Fe and Si tend to form intermetallic precipitates. The larger of these second-phase particles were found to be iron-rich, as revealed by energy-dispersive spectroscopy during our initial characterization of the material, consistent with previous reports in the literature \cite{Gao_defectsAA5252_2020}.

The cylindrical sample was affixed to an alumina rod with a ceramic adhesive and cured at room temperature for $>$\,$\SI{2}{\hour}$. The sample was then mounted on the rotation stage in the experimental hutch at beamline ID11 and recrystallized during two successive anneals of $\SI{30}{\minute}$ at $\SI{400}{\celsius}$ in a custom-built vertical furnace \cite{Furnace1999}. Subsequently, the specimen was annealed in $\SI[number-unit-product=\text{-}]{15}{\minute}$ intervals at $\SI{500}{\celsius}$ for a total of $\SI{90}{\minute}$. The complete annealing and measurement protocol is depicted in supplementary Fig.~\ref{fig:append:protocol}. After each annealing step, the sample was cooled to room temperature and characterized by DCT and PCT. All measurements were performed with a monochromatic x-ray box beam at an energy of $\SI{43.66}{\kilo\eV}$.

For DCT scans, diffraction spots were recorded using a FReLoN16 camera positioned $\SI{8.3}{\milli\meter}$ from the specimen, with an effective pixel side length of $\SI{1.78}{\micro\meter}$. To reduce the number of simultaneously diffracting grains and thereby avoid excessive spot overlap on the DCT detector, we adjusted an aperture to limit the incident beam dimensions to $\SI{900}{\micro\meter}\times\SI{380}{\micro\meter}$ (width\,$\times$\,height). After each annealing step, up to five adjacent scans were captured at vertical offsets of $\SI{350}{\micro\meter}$, resulting in an overlap of $\SI{30}{\micro\meter}$ between scans. Each scan, hereafter referred to as a DCT layer, covered a full $\SI{360}{\degree}$ rotation at a constant angular step of $\SI{0.05}{\degree}$, yielding 7200 diffraction images and a measurement time of about $\SI{25}{\minute}$ per DCT layer.

Because PCT does not require grain indexing and is therefore not limited by the number of grains in the irradiated volume, we increased the incident beam size to $\SI{1300}{\micro\meter}\times\SI{1300}{\micro\meter}$. Two PCT scans with a vertical overlap of $\SI{300}{\micro\meter}$ were recorded at each annealing step. Images were acquired every $\SI{0.2}{\degree}$ of rotation for a total of 1800 projections and a scan time of $\SI{5}{\minute}$ per layer. The images were captured by an Andor Marana camera with an effective pixel side length of $\SI{0.61}{\micro\meter}$, positioned $\SI{211.5}{\milli\meter}$ from the specimen.

\subsection{Reconstruction of diffraction-contrast datasets}
3D grain maps were reconstructed from the diffraction images with the DCT software package developed at ESRF based on Refs.~\cite{ludwig_three-dimensional_2009, reischig_advances_2013}. Modifications had to be made, however, because the bimodal nature of the microstructure introduced unique challenges. Our specific reconstruction routine had to accommodate large abnormal grains as well as thousands of smaller matrix grains.

In DCT, each diffraction spot corresponds to a 2D projection of the diffracting grain at a given sample rotation angle. Owing to the short sample-to-detector distance and the parallel beam optics, the size of each diffraction spot is directly related to the size of the diffracting grain \cite{ludwig_x-ray_2008}.
Consequently, abnormally large grains produce very large diffraction spots on the detector, as exemplarily shown in supplementary Fig.~\ref{fig:append:diffspots}. These are often truncated by the beam stop, or they may extend past the outer edges of the detector, resulting in inhomogeneous intensity distributions in the reconstructed grain volumes. These inhomogeneous intensity distributions make accurate segmentation challenging and lead to holes and overlaps within and around the abnormal grains in the assembled grain map.

To compensate for the effects of diffraction spot truncation, we modified the reconstruction and segmentation procedures in two ways. 
First, diffraction spots that had been automatically discarded during the initial reconstruction because of truncation were re-evaluated by visual inspection. We manually included spots that were only moderately truncated and did not overlap with neighboring diffraction spots in the reconstruction. Including these additional diffraction spots provided additional projections and improved the reconstruction of the abnormal grains. 
Second, the segmentation threshold was reduced to account for the presence of low-intensity regions within the reconstructed abnormal grains. Because diffraction spot truncation reduces the reconstructed intensity, particularly near grain perimeters, a higher threshold would have left these low-intensity regions unassigned. By adding projections and lowering the segmentation threshold, we obtained a more complete mapping of the abnormal grains, reducing both internal holes and gaps at the interface with surrounding matrix grains. The matrix grains, which did not suffer from truncation, were segmented with higher intensity thresholds. Overlapping regions between abnormal and matrix grains were then resolved by assigning disputed voxels to the grain with the higher reconstructed intensity. This criterion is applicable because the reconstructed intensities were normalized by grain volume, resulting in comparable intensity levels for abnormal and matrix grains.

The large number of matrix grains within the illuminated sample volume results in numerous overlapping diffraction spots. This complicates the process of assigning them to unique grains and ultimately leads to ambiguities in the reconstructed grain maps, where multiple grains can extend over the same voxels. To resolve such ambiguities, we filtered the matrix grains based on various characteristics, including completeness and solidity, with the latter defined as the ratio of the grain volume to the volume of its convex hull and calculated using the \code{regionprops3} function within the Image Processing Toolbox of MATLAB. Completeness denotes the ratio of the number of experimentally detected diffraction spots arising from a reconstructed grain to the total number of diffraction spots calculated for that grain by forward projection \cite{reischig_advances_2013, Fang2025}. A completeness of 1 indicates that all predicted diffraction spots were observed experimentally. A size-dependent threshold was applied to account for the generally lower completeness of smaller grains. Grains below a size-dependent completeness threshold (0.8 for the largest matrix grains, decreasing to 0.4 for the smallest) were excluded from the final grain map. Any remaining overlaps (disputed voxels) were then assigned to the grain with the higher reconstructed intensity at these positions in the grain map.

\subsection{Reconstruction and segmentation of phase-contrast datasets}
The projection images collected during the PCT scans were reconstructed on the Networked Interactive Computing Environment at ESRF using the tomography processing software Nabu \cite{paleo_nabu_2024} and Tomwer \cite{payno_tomwer_2023}. This workflow included background (flat-field) correction and phase retrieval using the Paganin algorithm \cite{paganin_simultaneous_2002}. The reconstruction was performed slice-wise, yielding a series of two-dimensional grayscale $x$-$y$ slices. The resulting stack of slices constitutes the reconstructed grayscale volume. These reconstructed volumes exhibited a pronounced low-intensity artifact that appeared as a ring-like structure in the individual $x$-$y$ slices (oriented perpendicular to the sample rotation axis), with both the intensity of the artifact and its appearance varying along the $z$-direction. To reduce the effect of this artifact on subsequent segmentation, the grayscale intensity of each $x$-$y$ reconstruction slice was normalized using the MATLAB function \code{imflatfield}.

Despite this normalization, residual intensity inhomogeneities remained, making it infeasible to apply a single intensity threshold to segment all of the small, bright, second-phase particles. Instead, we employed the image-processing software Dragonfly \cite{dragonfly_2022} to leverage its machine learning-based segmentation capabilities. To that end, we trained a convolutional neural network (CNN):\ specifically, a UNet++ architecture with a depth level of 5 and 32 filters in the first network layer, iterated for 100 epochs using five representative $x$-$y$ slices of the PCT reconstruction. In this procedure, several particles were manually selected from regions with different local grayscale intensities in the selected $x$-$y$-slices using the adaptive gaussian brush tool within Dragonfly. The segmentation targeted the upper intensity range using a constant sigma of 2 pixels and an intensity offset of 4 arbitrary units. To optimize local thresholding, the brush diameter---defining the localized neighborhood for background calculation---was dynamically adjusted between $\SI{3}{\micro\meter}$ and $\SI{14}{\micro\meter}$. Pixels located within the marked areas were assigned to class 1, while all remaining pixels were automatically assigned to class 2. (The network requires two classes for training.) The CNN was then applied to the same $x$-$y$ slices to automatically identify the particles. The resulting labeled images were visually examined, and additional particles that were not identified by the network were manually annotated. These additional particles were incorporated into the training data, after which the CNN was retrained and reapplied to the selected slices.
This iterative process was repeated until visual inspection indicated that the network reliably identified the majority of particles. 
The final trained CNN was subsequently applied to all $x$-$y$ slices of the PCT reconstruction, yielding a three-dimensional segmentation of the second-phase particles. The same segmentation procedure was then applied to the complete time series of reconstructed PCT scans.

\subsection{Registration of datasets}
Because the DCT and PCT scans were captured with different detectors, the reconstructed grain maps and phase-contrast maps (including the segmented particles) were initially not aligned. As part of each DCT measurement, the DCT detector was used to record a complete projection series of low-quality absorption scans at the same effective pixel size as DCT. The resulting tomographic reconstruction contains many of the high-contrast features (such as edges, larger second-phase particles, etc.) that are also visible in the corresponding phase-contrast map. After identifying matching slices in both datasets, we downscaled the phase-contrast map from a voxel side length of $\SI{0.61}{\micro\meter}$ to $\SI{1.78}{\micro\meter}$.

Next, we registered the downscaled volume to the absorption volume (measured with the DCT detector) using a rigid-body transformation determined by the MATLAB function \code{imregtform}. We then applied the transformation to the centroids of the higher-resolution segmented particle map using the MATLAB function \code{transformPointsForward}, allowing them to be mapped onto the DCT grain map.

In the as-reconstructed state, the DCT grain maps recorded at different annealing times were slightly misaligned, owing to slight displacements of the measurement setup (which likely occurred when the detectors were moved into and out of position during each measurement sequence). A pairwise registration of datasets from adjacent annealing intervals was performed by estimating and applying a rigid-body transformation to the extracted GB networks \cite{dake_experimental_2020}, starting with the grain map at $\SI{15}{\minute}$. The same transformation matrices were then applied to the particle maps to ensure alignment between the grain and particle maps over the entire duration of annealing.

\subsection{Grain tracking}
After registration, a tracking algorithm \cite{dake_2016, dake_experimental_2020} was applied to the grain maps to ensure that grains have the same labels across all time steps. The algorithm uses two criteria to find matching grains:\ first, the fractional volume overlap and, second, the misorientation between pairs of grains in the current and previous time steps, tracking backward from later to earlier times. Grains with a minimum volume overlap of 25\% and a maximum misorientation of $\SI{0.5}{\degree}$ in two adjacent time steps are stored as a match. If there are multiple matches for a given grain at the later time step, the pair with the largest volume overlap is stored as the final match. Using this procedure, 94\% to 97\% of all grains in a later dataset were successfully found in its immediately preceding dataset.

\section{Results}
Initial reconstructions revealed two features that greatly simplified the subsequent analysis. First, we noticed that all abnormal grains were contained within a single DCT layer. To increase the chances of detecting an abnormal grain, we had chosen to map the largest volume practically possible given the time constraints of the experiment, which was five adjacent layers of DCT mapping. All abnormal grains turned out be located within the first DCT layer. The second observation concerned the abundance of matrix grains in this particular layer. Directly after the recrystallization protocol (1\,h at $\SI{400}{\celsius}$), the first DCT layer contained more than 30\,000 grains. Following the initial 15-min anneal at $\SI{500}{\celsius}$, however, the number of grains dropped to about 9\,000. Because the computational time for data reconstruction and segmentation increases significantly with the number of grains, we chose to begin the tracking of grain growth from $t = \SI{15}{\minute}$ rather than $t = \SI{0}{\minute}$.  These reductions in the spatial and temporal range of the data drastically reduced the time and resources required for data analysis, while still allowing for the growth dynamics of both abnormal and matrix grains to be captured.

\subsection{Evolution of 3D grain microstructure}
\begin{figure}
    \captionsetup{width=\textwidth}
    \centering
    \includegraphics[scale=0.75]{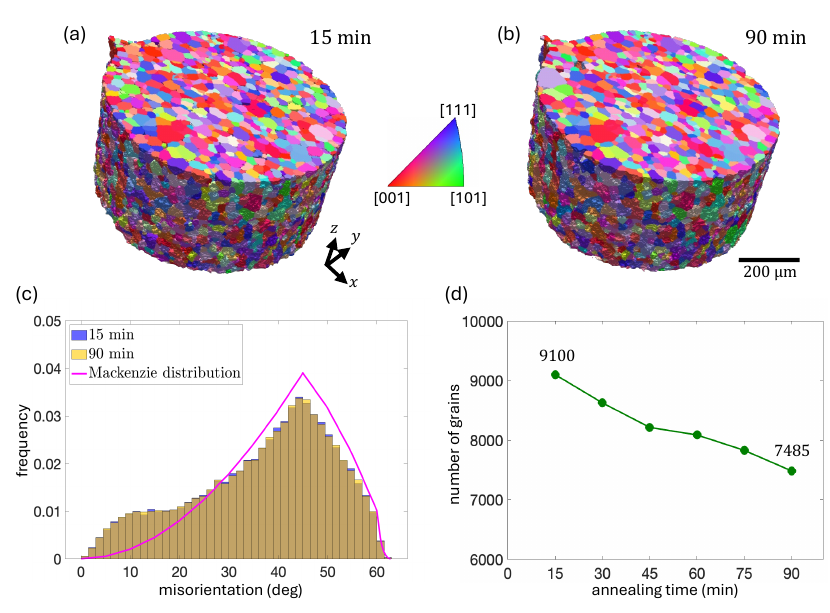}
    \caption{Evolution of 3D grain microstructure in AA5252 at $\SI{500}{\celsius}$. Reconstructed grain maps for the selected sample layer following annealing for (a)~$\SI{15}{\minute}$ and (b)~$\SI{90}{\minute}$. Each grain is colored based on its crystal orientation according to the inverse pole figure (IPF) color coding scheme relative to the sample $z$-axis (see inset). The grain maps were visualized using ParaView \cite{ParaView}. (c)~The lattice misorientation distribution was determined between all nearest-neighbor grains after $\SI{15}{\minute}$ and $\SI{90}{\minute}$. For comparison, the Mackenzie distribution \cite{mackenzie_second_1958} for randomly oriented cubic crystals is shown (pink curve). (d)~Change in the total number of grains in the segmented layer.}
    \label{fig:3D_grainmaps}
\end{figure}

Six DCT grain maps were reconstructed, spanning the $\SI{500}{\celsius}$ anneal from $\SI{15}{\minute}$ to $\SI{90}{\minute}$ in $\SI[number-unit-product=\text{-}]{15}{\minute}$ intervals. Fig.~\ref{fig:3D_grainmaps} summarizes the evolution of the reconstructed layer by combining 3D views of the first and last grain maps with the corresponding Mackenzie distributions and total grain count. Although the outer surfaces of the specimen suggest only modest grain growth, the internal evolution was substantial, especially in the neighborhood of two abnormally growing grains. Nearly half of the grains that vanished (from 9\,100 to 7\,485) were consumed by the two abnormal grains.

\begin{figure}
      \captionsetup{width=\textwidth}
    \centering
        \includegraphics[scale=0.8]{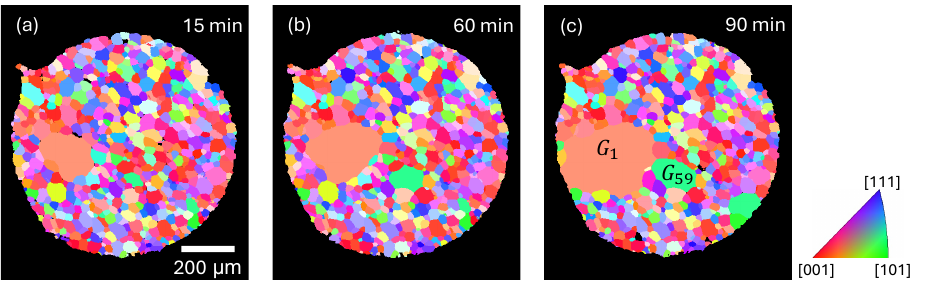}
    \caption{Identification of abnormal grains in AA5252. Cross sections (at $z \approx \SI{75}{\micro\meter}$) through the 3D grain maps after (a) $\SI{15}{\minute}$, (b) $\SI{60}{\minute}$, and (c) $\SI{90}{\minute}$ of annealing at $\SI{500}{\celsius}$ reveal the presence of at least two abnormally growing grains, labeled $G_1$ (orange) and $G_{59}$ (green). Pixels in the sample interior that were not assigned to a grain during reconstruction are colored black.}
    \label{fig:cross_sections}
\end{figure}

To help visualize the internal grain evolution, Fig.~\ref{fig:cross_sections} presents 2D cross sections through the reconstructed volume at $z \approx \SI{75}{\micro\meter}$ for selected annealing times. In contrast to the relatively modest changes visible at the specimen surface, these cross sections reveal pronounced abnormal grain growth within the interior. The orange grain, $G_1$, is already substantially larger than the surrounding matrix grains at the $\SI[quantity-product=\text{-}]{15}{\minute}$ mark, and $G_1$ maintains this growth advantage over its neighbors throughout the annealing treatment. A second abnormal grain, $G_{59}$, enters the cross section at $\SI{60}{\minute}$, also exhibiting abnormally fast growth before impinging on $G_1$ by $\SI{90}{\minute}$.

\begin{figure}
      \captionsetup{width=\textwidth}
    \centering
    \includegraphics[scale=0.8]{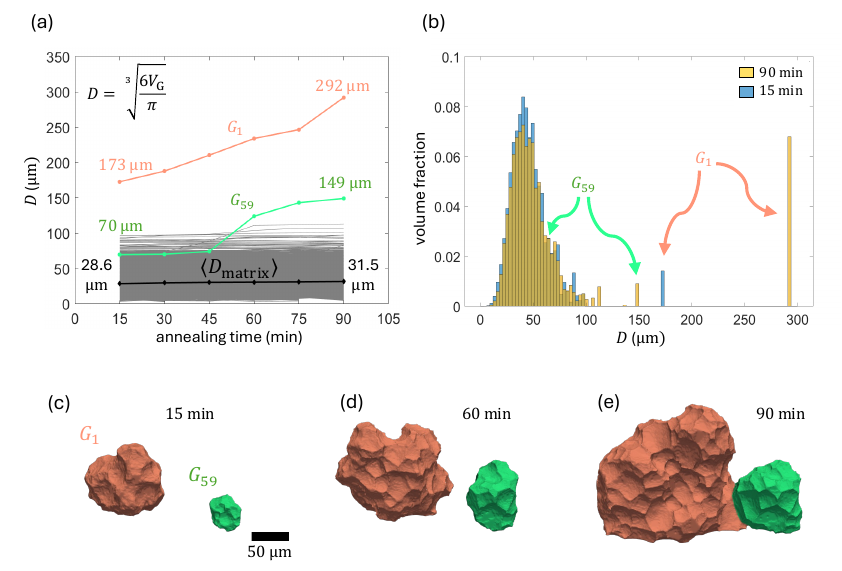}
        \caption{Evolution of abnormal grains in AA5252 at $\SI{500}{\celsius}$. (a)~Evolution of the volume-equivalent sphere diameter $D$, calculated from the grain volume $V_\mathrm{G}$, for abnormally growing grains (orange and green lines) in comparison to a subset of the matrix grains (gray lines). (b)~Volume-weighted grain size distributions (GSDs) at $\SI{15}{\minute}$ and $\SI{90}{\minute}$, with arrows marking the positions of grains $G_1$ and $G_{59}$ within these distributions. After $\SI{90}{\minute}$ of annealing,  both grains lie well outside the matrix GSD. 3D visualization of the abnormally growing grains at time steps (c)~$\SI{15}{\minute}$, (d)~$\SI{60}{\minute}$ and (e)~$\SI{90}{\minute}$.}
    \label{fig:3D_G1_G59}
\end{figure}

The abnormal evolution of grains $G_1$ and $G_{59}$ becomes more striking once it is compared to the growth of the remaining grains in the reconstructed volume. Using the equivalent-sphere diameter as a 1D measure for grain size, we can compare the growth trajectories of the abnormal grains to those of the ``normal'' matrix grains. These trajectories are plotted in Fig.~\ref{fig:3D_G1_G59} along with the grain size distributions after $\SI{15}{\minute}$ and $\SI{90}{\minute}$ of annealing. Over the entire $\SI[quantity-product=\text{-}]{75}{\minute}$ duration at $\SI{500}{\celsius}$, the matrix grains remain remarkably stable, increasing only marginally in size from a mean diameter of $\SI{28.6}{\micro\meter}$ to $\SI{31.5}{\micro\meter}$. At the start of our analysis, grain $G_1$ already lies beyond the upper range of the unimodal distribution of matrix grain sizes, and $G_1$ grows at a much-faster-than-average pace throughout the experiment from $\SI{173}{\micro\m}$ to $\SI{292}{\micro\m}$. Grain $G_{59}$, on the other hand, starts out well within the distribution of matrix grain sizes at $\SI{15}{\minute}$, where it remains for the next $\SI{30}{\minute}$, before rapidly increasing in size until eventually impinging upon $G_1$. Apparently, in this single dataset, we not only succeeded in capturing the persistent growth of two abnormal grains but also the emergence of one of them.

\subsection{Distribution and stability of pinning particles}
\begin{figure}
    \captionsetup{width=\textwidth}
    \centering
    \includegraphics[scale=0.8]{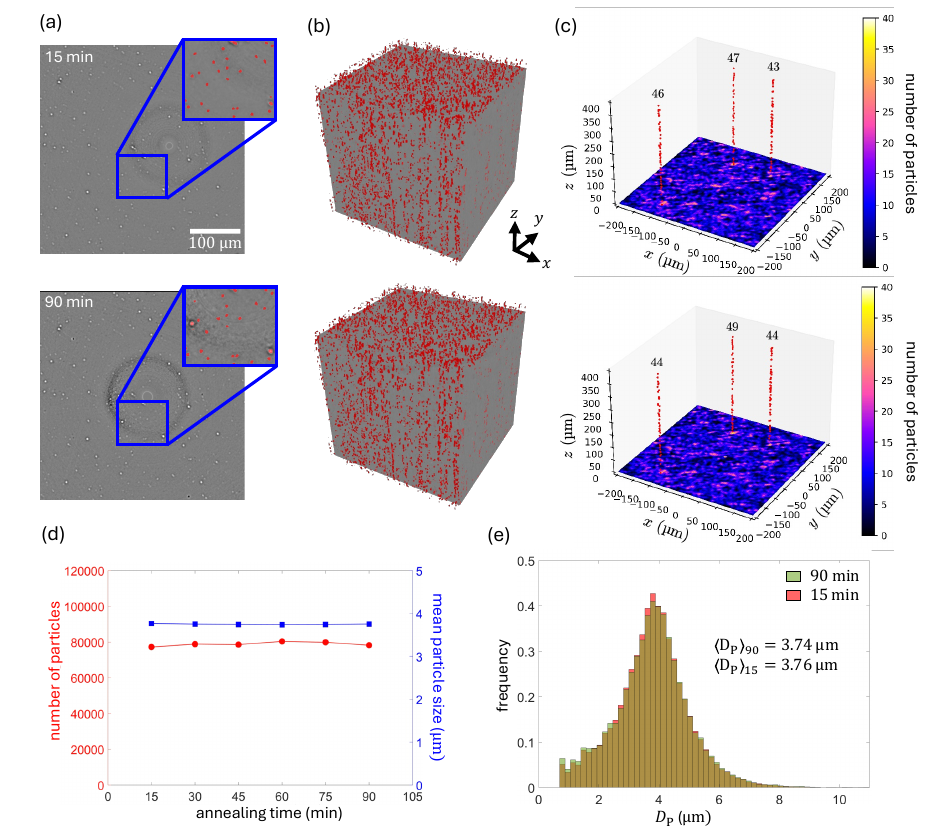}
    \caption{Spatial arrangement and evolution of second-phase particles in AA5252 at $\SI{500}{\celsius}$. (a)~Central region of the reconstructed phase-contrast maps at (top) $\SI{15}{\minute}$ and (bottom) $\SI{90}{\minute}$. Second-phase particles (high-contrast regions) located within the blue squares are marked in red following segmentation by Dragonfly (insets). (b)~Reconstructed phase-contrast volumes of the central sample region for the same two time steps, with segmented particles marked in red. (c)~Particle clustering within the volume shown in (b) is quantified by binning particle center-of-mass positions projected onto the $x$–$y-$plane into $4\times4$ pixel bins and counting the particles per bin. The particles belonging to the three bins with the highest particle counts are plotted in 3D, revealing a tendency for alignment along the $z$-axis. Further details on the influence of the spatial bin width on the observed particle clustering effect are provided in supplementary Fig.~\ref{fig:append:particles_2}. Quantifying the size of segmented particles within the entire layer in terms of an equivalent-sphere diameter, we obtain (d)~the~total number of particles and mean particle size for all annealing steps as well as (e)~their size distributions at $\SI{15}{\minute}$ and $\SI{90}{\minute}$.}
    \label{fig:particles_1}
\end{figure}

While DCT provides information on the evolving grain structure, high-resolution PCT reveals the distribution of second-phase particles in the irradiated volume. Segmentation of the reconstructed PCT volumes yielded particle maps containing approximately 80\,000 particles, with the smallest segmented particles measuring $\SI{0.76}{\micro\meter}$ in diameter, close to the effective voxel side length of $\SI{0.61}{\micro\meter}$. The resulting particle count is roughly an order of magnitude greater than the number of grains identified within the same analyzed volume. This relatively high ratio indicates that a substantial number of potential pinning particles may interact with the grain boundary network. Three-dimensional visualization of the segmented particles (Fig.~\ref{fig:particles_1}) reveals that they are not uniformly distributed throughout the volume but, rather, tend to align along the $z$-direction, forming particle-rich columns or stringers. This morphology is perhaps not surprising, given that the $z$-axis of the specimen coincides with the rolling direction of the sheet from which it was cut.

The second-phase particles found by PCT are remarkably stable over the course of the experiment. Not only does the total number of particles detected remain nearly constant (within the uncertainty of the segmentation routine), but the mean size and spatial arrangement of the particles are also essentially unchanged. This conclusion is bolstered by direct comparison of the segmented particle maps at different annealing times. Cross sections through the same location in the specimen, taken from the beginning and end of the experiment, reveal second-phase particles at the exact same positions and of the same size, and their columnar arrangement likewise remains unchanged. No evidence of particle coarsening or measurable particle motion was observed over the 90-minute duration of the experiment.

The pinning effect exerted by this static particle dispersion on grain growth can be estimated by calculating the Zener-limited grain size, $D_\mathrm{lim}$, which is given by
\begin{equation}
    D_\mathrm{lim} = k \frac{d}{(f_V)^m},
    \label{eq:Rlim}
\end{equation}
where $f_V$ denotes the particle volume fraction, $d$ the mean particle diameter, $k$ a proportionality constant, and $m$ an empirical exponent \cite{manohar_five_1998}.
Following Manohar \textit{et al.} \cite{manohar_five_1998}, we use $m=1$ and $k=0.17$, as these values are commonly reported for particle volume fractions below 5\%. Substituting the experimentally determined mean particle diameter of $\SI{3.75}{\micro\meter}$ and the particle volume fraction of 1.6\% into Eq.~(\ref{eq:Rlim}), we calculate a Zener-limited grain size of $D_\mathrm{lim}\approx\SI{40}{\micro\meter}$. Although this estimate exceeds the measured mean matrix grain size ($\left<D_\mathrm{matrix}\right> \approx \SI{30}{\micro\meter}$), the two values are of the same order of magnitude. Moreover, because the smallest second-phase particles are the most difficult to segment reliably, the measured particle volume fraction is likely underestimated, while the mean particle size is correspondingly overestimated. Insertion of more accurate values into Eq.~(\ref{eq:Rlim}) would decrease $D_\mathrm{lim}$ accordingly, indicating that the calculated value represents an upper bound. Thus, within the context of Zener pinning the measured dispersion of second-phase particles can adequately account for the observed stagnation of matrix grains during the experiment. 

\subsection{Grain boundary migration at abnormal–matrix interfaces}
\begin{figure}
    \captionsetup{width=\textwidth}
    \centering
    \includegraphics[scale=0.8]{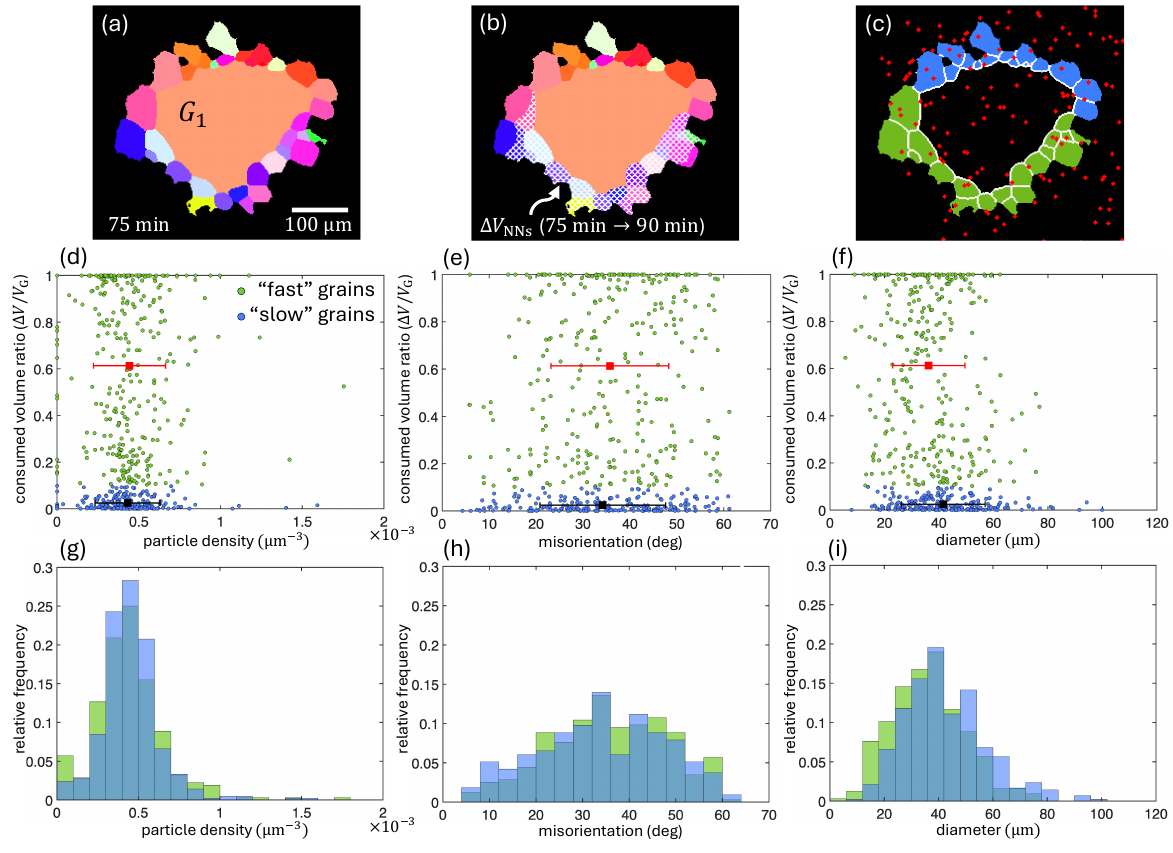}
    \caption{Classification and comparison of nearest-neighbor grains consumed rapidly or slowly by abnormal grain $G_1$. A cross section through $G_1$ and its neighboring grains is shown in (a) at the $\SI{75}{\minute}$ annealing interval, while (b) highlights with white hatching the regions of neighboring grains consumed by $G_1$ during the subsequent $\SI[quantity-product=\text{-}]{15}{\minute}$ anneal. The neighboring grains are classified in (c) according to their relative volume loss:\ grains that lose more than 10\% of their volume to $G_1$ during an annealing interval are termed ``fast'' grains (green), while those that lose 10\% or less are termed ``slow'' grains (blue). Second-phase particles in the local environment of $G_1$ are shown in red. For five annealing intervals, the relative volume consumed, $\Delta V/V_\mathrm{G}$, is plotted against (d) particle density, (e) misorientation and (f) grain diameter for the fast and slow populations. The corresponding mean values and standard deviations are indicated in red for the fast grains and black for the slow grains. Binning the data yields relative-frequency histograms, allowing direct comparison of the (g) particle-density, (h) misorientation and (i) grain diameter distributions of the two populations.}
    \label{fig:fast_slow_NNs}
\end{figure}

While second-phase particles effectively pin the matrix grains, substantial migration still occurs at boundaries separating the abnormal grains from the surrounding matrix. However, not all of these GBs migrate to the same extent:\ some of the neighboring grains are consumed rapidly by an abnormal grain, while others persist over several annealing intervals. To identify characteristics of the local microstructure that may be associated with these differences, we examine the neighboring grains of abnormal grains at each time step and group them according to how rapidly they are consumed. This classification is illustrated for abnormal grain $G_1$ with the help of 2D cross sections in Fig.~\ref{fig:fast_slow_NNs}. For brevity, neighboring grains that lose more than 10\% of their volume to $G_1$ during a given annealing interval are termed ``fast'' grains, whereas those that retain at least 90\% of their volume are termed ``slow'' grains.

After separating the neighboring grains of $G_1$ into these two populations, we can compare and contrast parameters such as particle density, misorientation and grain size. In Fig.~\ref{fig:fast_slow_NNs}\,(d)--(f), the relative volume consumed by the abnormal grain, $\Delta V / V_\mathrm{G}$, is plotted against each of these quantities. The 10\% cutoff separating the two populations is clearly visible, as is the large scatter in the data. Binning the corresponding values for the fast and slow grains yields the relative-frequency histograms shown in the last three panels (Fig.~\ref{fig:fast_slow_NNs}\,(g)--(i)), facilitating direct comparison of the two populations. The particle-density distributions are quite similar for both populations; however, slight differences are apparent in the misorientation and, more prominently, in the grain size distributions. Neighboring grains that are consumed quickly are, on average, more than $\SI{5}{\micro\meter}$ smaller than those that are consumed slowly.

To determine whether the differences between these two populations are statistically significant, we first apply Welch's t-test to compare their mean values. At a significance level of $\alpha = 0.01$, the resulting $p$-values for particle density ($p=0.60$) and misorientation ($p=0.015$) exceed the significance threshold, whereas the difference in grain size is highly significant ($p=7.2\times10^{-11}$). The corresponding 99\% confidence interval for the difference in mean grain diameter is approximately \qtyrange[range-units=single]{4.3}{9.7}{\micro\m}. However, differences in the shape of a distribution, particularly near its extremes, may not be reflected strongly in the mean value. This is especially relevant for misorientation distributions, which are bounded and non-Gaussian. We therefore also apply a two-sample Kolmogorov-Smirnov (KS) test, which compares the full distributions rather than only their mean values. At the same 1\% significance level, the KS test provides no evidence for a significant difference between the particle-density distributions ($p=0.13$), but does indicate differences in the misorientation ($p=0.008$) and grain size distributions ($p=1.6\times10^{-5}$). Applying the same analysis to the neighboring grains of the second abnormal grain ($G_{59}$) yields qualitatively similar results, providing a second case in which grain size and, to a lesser extent, crystallographic misorientation distinguish the two populations of nearest neighbor grains.

Contrary to what one might naively expect and to what is often hypothesized in the literature, the local particle density does not distinguish between regions that are consumed rapidly and those that are consumed more slowly. For $G_1$, neither the mean particle density nor the overall particle-density distribution differs significantly between the populations of ``fast'' and ``slow'' grains. The analysis of $G_{59}$ leads to the same general conclusion; in fact, for $G_{59}$ the particle density turns out to be slightly higher for the neighboring grains that are consumed more rapidly! Moreover, the particle densities measured directly at the migrating interfaces after each 15-min annealing interval are comparable for the two groups. Taken together, these observations provide no indication that the abnormal grains preferentially grow into regions of reduced particle density. Instead, grain size is the clearest characteristic differentiating between rapidly and slowly consumed neighboring grains, accompanied by a weaker disparity in GB misorientation, with a slightly larger fraction of low-angle GBs separating the abnormal grains from their more slowly consumed neighbors.

\section{Discussion}
The suppression of normal grain growth is a prerequisite for abnormal grain growth. In the aluminum alloy AA5252, this is apparently achieved by second-phase particles that cause growth to stagnate at a mean grain size of approximately $\SI{30}{\micro\m}$. These pinning particles are also remarkably stable---even at the high homologous annealing temperature of $\SI{500}{\celsius}$, they neither coarsen nor dissolve, and although the particles create the stagnant matrix required for AGG, local variations in particle density do not appear to determine where abnormal growth proceeds. Thus, the central question becomes how boundaries of some grains are able to overcome essentially the same pinning conditions that arrest the surrounding matrix grains. More specifically, and in the sense of our investigation, we can ask:\ how does a matrix grain first escape the pinned population to emerge as an abnormal grain, and, once it has done so, what conditions facilitate the persistence of accelerated growth?

\begin{figure}
      \captionsetup{width=\textwidth}
    \centering
    \includegraphics[scale=1]{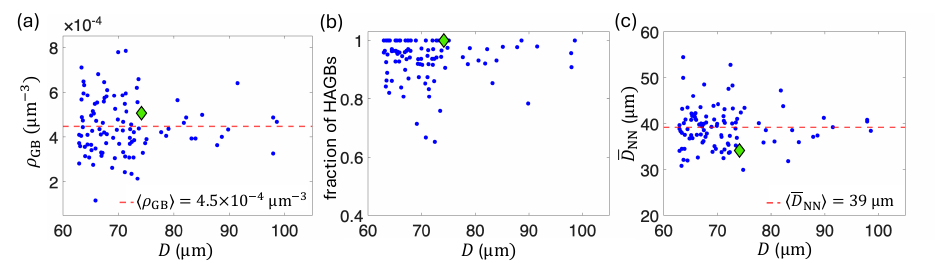}
    \caption{Comparison of grain $G_{59}$'s local environment at the $\SI[quantity-product=\text{-}]{45}{\minute}$ annealing mark to that of the other 100 largest (non-surface) grains in the specimen. Panel (a) compares the particle number density, $\rho_\mathrm{GB}$, determined in a two-voxel shell around each grain surface with the volume-equivalent sphere diameter $D$. Grain $G_{59}$, marked by the green diamond, has a particle density above the average of the comparison grains. The fraction of HAGBs is plotted for the same grains in (b), revealing that the neighbors of $G_{59}$ all have misorientations greater than $\SI{15}{\degree}$. In (c), the local grain size environment is quantified by the mean nearest-neighbor diameter $\overline{D}_\mathrm{NN}$. Among the characteristics examined, the emergence of $G_{59}$ is associated with an exclusively high-angle GB network and smaller-than-average neighboring grains, rather than with a locally reduced particle density.}
    \label{fig:emergence_G59_45min}
\end{figure}

Grain $G_{59}$ provides a rare opportunity to address the question of emergence. Unlike $G_1$, which was already substantially larger than the matrix grains at the beginning of the annealing experiment, $G_{59}$ initially lies within the size range of matrix grains and begins to grow abnormally only after $\SI{45}{\minute}$ of annealing. Although $G_{59}$ is larger than the mean matrix grain from the very beginning of the experiment, its size alone cannot explain its anomalous growth behavior, since many grains of comparable or even greater size do not undergo a transition to abnormal growth. The comparison in Fig.~\ref{fig:emergence_G59_45min} with the 100 largest non-surface grains that remain part of the matrix provides a means of identifying other characteristics that may have favored the emergence of $G_{59}$. (Only non-surface grains were considered in this comparison to minimize potential biases arising from reduced particle segmentation near specimen boundaries and from the different GB environments of grains intersecting a free surface.) Of the parameters examined, $G_{59}$ is distinguished by an exclusively high-angle GB network and, perhaps most notably, by a small nearest-neighbor grain size. In contrast, the particle density at its surface is above rather than below the average of the comparison grains. Evidently, neither a locally reduced particle density nor particle coarsening or dissolution appears necessary for the emergence observed here. This differs from the behavior reported by Lu \textit{et al.}~\cite{lu_dynamics_2020}, who associated abnormal growth in an Al-Cu alloy with relatively particle-poor regions. In the present case, a favorable combination of local grain geometry and GB properties appears to provide a more plausible explanation. However, since only a single emergence event is captured, these characteristics should be regarded as conditions associated with the emergence of $G_{59}$ rather than as evidence of a unique triggering mechanism.

The importance of the local grain-size relationship is reinforced by the subsequent growth of both abnormal grains. Among the microstructural characteristics considered in the nearest-neighbor analysis, grain size most clearly distinguishes grains that are consumed rapidly from those that are consumed more slowly (as shown in Fig.~\ref{fig:fast_slow_NNs}). GB misorientation provides a weaker distinction, whereas local particle density provides essentially none. The significance of this result lies not simply in the preferential consumption of smaller grains, but in what grain size represents. Differences in grain size are associated with differences in the geometry and curvature of the surrounding GB network and, consequently, with differences in capillary driving pressure. The dependence of growth behavior on neighboring grain size therefore points toward variations in capillary driving force, rather than variations in local particle density, as an important factor governing continued abnormal grain boundary migration. The weaker dependence on misorientation further suggests that GB character modulates this response, since a sufficiently large curvature need not result in rapid migration if the energy or mobility of a particular GB is low.

\begin{figure}
      \captionsetup{width=\textwidth}
    \centering
    \includegraphics[scale=0.7]{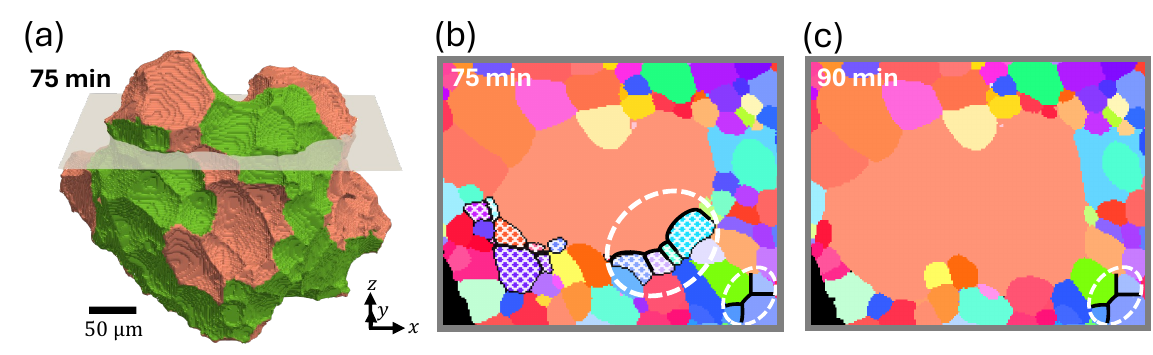}
   \caption{Proposed mechanism for persistent abnormal grain growth. A 3D visualization of grain $G_1$ after $\SI{75}{\minute}$ is presented in (a), with GBs shared with nearest-neighbor grains that lose more than 75\% of their volume to $G_1$ during the subsequent $\SI{15}{\minute}$ marked in green. The gray plane indicates the position of the cross sections in (b) and (c). A qualitative comparison in (b) reveals higher curvature along several abnormal/matrix GBs than along the surrounding matrix/matrix GBs, particularly in the encircled regions. Neighboring grains that are subsequently consumed by more than 75\% are indicated by hatching. After a further $\SI{15}{\minute}$ of annealing, $G_1$ has consumed large portions of these grains. Meanwhile, the surrounding matrix/matrix GBs, including the flat ones in the lower-right encircled region, remain largely immobile (c). The evolving geometry illustrates how rapid growth of some grains can generate strongly curved abnormal/matrix GBs and thereby promote further migration.}
    \label{fig:possible_mechanism}
\end{figure}

This interpretation can be understood more directly by considering the evolving geometry of the abnormal/matrix GBs, as illustrated qualitatively in Fig.~\ref{fig:possible_mechanism} for grain $G_1$.  When grains of similar sizes intersect, they tend to form relatively flat GBs; however, when a large grain forms GBs with many smaller neighbors, the boundaries are not only concave outward, but their curvatures also depend on the grain size ratio. Although the exact value of the boundary curvature depends on the specific configuration of neighboring grains and on their boundary character, in general, the directional force-balance required at grain boundary triple junctions \cite{Herring1951a,Herring1951b} leads to higher curvatures as the abnormal-to-matrix size ratio increases. Thus, it makes physical sense for smaller grains to be consumed more rapidly. As this happens, however, the remaining neighbors become increasingly peninsula-like or even isolated within the advancing abnormal grain, producing more strongly curved GB segments even where interfaces were initially comparatively flat. The associated increase in driving pressure may thereby allow further GBs to overcome particle pinning. 

The migration of boundary segments necessarily alters the geometry---and, thus, the driving pressure---of neighboring segments, which can, at the abnormal/matrix interface, provide additional pathways for abnormal growth to continue. Curvature alone, though, is insufficient to determine whether a GB migrates. For example, the strongly curved LAGB associated with the yellow-beige grain above~$G_1$ in Fig.~\ref{fig:possible_mechanism}\,(b,c) remains nearly stationary over several annealing intervals. Whether or not a specific GB unpins and migrates also depends on GB-specific properties. Persistent abnormal growth can therefore be viewed as the result of a continuously evolving competition between capillary driving pressure and particle pinning, with GB character determining how readily a particular boundary responds.

Viewed at the grain scale, this local picture is consistent with the mechanism proposed by Humphreys for persistent AGG in particle-pinned microstructures \cite{humphreys_unified_1997_2}. In his framework, a sufficiently large grain can continue to grow provided its size relative to the surrounding matrix remains above a threshold value set by the pinning-particle conditions. The preferential consumption of smaller nearest-neighbor grains that we observed for both $G_1$ and $G_{59}$ is consistent with this model and, in the context of curvature, provides an interpretation of the size-ratio criterion. Importantly, for persistence of this mechanism, the nearly stagnant matrix generates positive feedback once an abnormal grain has gained a sufficient size advantage. Further growth increases the size of the abnormal grain while the characteristic matrix grain size changes only marginally, so that the size disparity responsible for the growth advantage increases rather than diminishes. At the same time, consumption of neighboring grains continuously exposes the abnormal grain to new matrix grains and modifies the local geometry of its GBs. Wherever the resulting capillary driving pressure is sufficient to overcome particle pinning, the corresponding GB will be sufficiently mobile, and migration can continue. In this way, the growth process itself reinforces the conditions that permit further abnormal growth.

If this picture of particle-meidated AGG is correct, the distinction drawn between the \textit{emergence} and \textit{persistence} of an abnormal grain may be primarily phenomenological rather than mechanistic. At the scale of an individual GB, there is no fundamental difference between a boundary segment belonging to a grain that has not yet been classified as abnormal and one belonging to an already established abnormal grain. In both cases, migration is governed by the same underlying competition between capillary driving pressure and pinning pressure, modulated by GB-specific properties. During emergence, favorable local geometry and GB character must first facilitate sustained migration despite the pinning conditions that arrest most of the matrix. Thermally activated fluctuations of GBs may allow individual boundary segments to sample their local energy landscape and overcome pinning barriers when local conditions are sufficiently favorable. Once GB migration has given rise to a meaningful size advantage, the nearly stagnant matrix reinforces the size advantage criterion, thereby enabling continued growth to occur. Although the present experiment cannot resolve the exact event(s) that initiated this transition, the observations narrow the mechanistic problem considerably:\ rather than requiring abnormal grains to encounter systematically weaker particle pinning, the emergence and persistent growth of abnormal grains can be understood in terms of how local grain geometries and GB properties allow individual boundary segments to overcome the pinning field that keeps the surrounding matrix nearly stagnant. In a subsequent manuscript, we will apply the methods described in Ref.~\cite{Wilhelm2026} to quantify the local GB curvature and migration kinetics at the abnormal/matrix interfaces of $G_1$ and $G_{59}$, thus enabling the balance between capillary driving pressure and particle pinning proposed here to be tested more directly.

\section{Conclusions}
By combining time-resolved diffraction-contrast and phase-contrast tomography, we investigated the emergence and persistent growth of abnormal grains in a commercial Al-Mg alloy (AA5252) containing finely dispersed second-phase particles. The principal findings are summarized as follows:

1. Second-phase particles suppress normal grain growth but do not prevent abnormal grain growth. The particle population remained essentially unchanged during annealing, and the matrix grains exhibited very little growth, consistent with Zener pinning. Nevertheless, two abnormal grains were able to grow substantially at the expense of the surrounding matrix.

2. Local variations in particle density do not appear to determine where abnormal grain growth occurs. Neither the emergence of an abnormal grain nor the preferential consumption of neighboring matrix grains was associated with locally reduced particle density. Thus, while particle pinning creates the conditions required for AGG, spatial variations in particle density do not appear to mediate its progression.

3. Local grain geometry and grain boundary properties play an important role in abnormal growth. Smaller neighboring grains were consumed preferentially, while grain boundary misorientation exhibited a weaker but still discernible influence. These observations suggest that variations in local curvature and the associated capillary driving pressure, modulated by grain boundary properties, determine which boundaries can overcome particle pinning and continue to migrate. The resulting changes in boundary geometry near small grains that vanish may further reinforce abnormal growth.

4. The emergence and persistence of abnormal grains may be governed by the same underlying mechanism. The observed emergence of abnormal grain $G_{59}$ was associated with relatively small neighboring grains and an exclusively high-angle grain boundary network, rather than with reduced particle pinning. These findings suggest that favorable local grain geometry and grain boundary properties can initiate abnormal growth, after which the increasing size advantage over the nearly stagnant matrix promotes continued AGG.

\section*{Acknowledgments}
This research is funded by the German Research Foundation (DFG) project number 453092613 (KR 1658/9-1,  SCHM 997/41-1). We thank Constellium and Baptiste Flipon (CEMEF, MINES ParisTech {\textbar} PSL University) for providing the cold-rolled sheet of AA5252, and we gratefully acknowledge ESRF for granting measurement time at beamline ID11 through proposal MA-5829. We would also like to thank Kyocera for providing the ceramic rods used to mount the samples at the beamline.

\bibliographystyle{unsrt}
\bibliography{bibliography.bib}

@article{Park_Fe_Si_2018,
    author = {Park, H. and Kim, D.-Y. and Hwang, N.-M. and Joo, Y.-C. and Han, C.H. and Kim, J.-K.},
    title = {Microstructural evidence of abnormal grain growth by solid-state wetting in Fe-3\%Si steel},
    journal = {Journal of Applied Physics},
    volume = {95},
    pages = {5515--5521},
    year = {2004}
}

@book{dunn_secondary_1966,
	title = {Recrystallization, Grain Growth and Textures},
	pages = {461--521},
	publisher = {American Society for Metals},
	author = {Dunn, C.G. and Walter, J.L.},
	year = {1966},
}

@article{humphreys_unified_1997_1,
title = {A unified theory of recovery, recrystallization and grain growth, based on the stability and growth of cellular microstructures—{I}. {T}he basic model},
journal = {Acta Materialia},
volume = {45},
pages = {4231-4240},
year = {1997},
author = {Humphreys, F.J.},
}

@article{poulsen_three-dimensional_2001,
	title = {Three-dimensional maps of grain boundaries and the stress state of individual grains in polycrystals and powders},
	volume = {34},
	pages = {751--756},
	journal = {Journal of Applied Crystallography},
	author = {Poulsen, H.F. and Nielsen, S.F. and Lauridsen, E.M. and Schmidt, S. and Suter, R.M. and Lienert, U. and Margulies, L. and Lorentzen, T. and Juul Jensen, D.},
    year={2001}
}

@article{sorensen_multigrain_2012,
	title = {Multigrain crystallography},
	volume = {227},
	pages = {63--78},
	journal = {Zeitschrift für Kristallographie},
	author = {Sørensen, H.O. and Schmidt, S. and Wright, J.P. and Vaughan, G.B.M. and Techert, S. and Garman, Elspeth F. and Oddershede, J. and Davaasambuu, J. and Paithankar, K.S. and Gundlach, C. and Poulsen, H.F.},
	year = {2012}
}

@article{liu_three-dimensional_2011,
	title = {Three-dimensional orientation mapping in the transmission electron microscope},
	volume = {332},
	pages = {833--834},
	journal = {Science},
	author = {Liu, H.H. and Schmidt, S. and Poulsen, H.F. and Godfrey, A. and Liu, Z.Q. and Sharon, J.A. and Huang, X.},
	year = {2011},
}

@article{king_first_2013,
	title = {First laboratory {X}-ray diffraction contrast tomography for grain mapping of polycrystals},
	volume = {46},
	pages = {1734--1740},
	journal = {Journal of Applied Crystallography},
	author = {King, A. and Reischig, P. and Adrien, J. and Ludwig, W.},
	year = {2013},
}

@article{zhu_3d_2023,
	title = {3{D} characterization of abnormal grain growth in nanocrystalline nickel},
	volume = {1},
	pages = {e19},
	journal = {Materials Genome Engineering Advances},
	author = {Zhu, W. and Huang, X. and Cai, W. and Huang, T. and Wu, G. and Huang, X.},
	year = {2023},
}

@article{wang_4d_2025,
	title = {4{D} {O}bservations of the initiation of abnormal grain growth in commercially pure {N}i},
	volume = {264},
	pages = {116715},
	journal = {Scripta Materialia},
	author = {Wang, Y. and Xu, Z. and Muralikrishnan, V. and Harley, J.B. and Tonks, M.R. and Rohrer, G.S. and Krause, A.R.},
	year = {2025}
}

@article{Ånes_2026,
title = {Time-resolved 4{D} characterisation of an abnormal grain growth event in pure iron},
publisher = {IOP Publishing},
volume = {1350},
pages = {012017},
journal = {IOP Conference Series: Materials Science and Engineering},
author = {Ånes, H. W. and Heinig, M. F. and Wang, P. and Oddershede, J. and Purohit, R. R. P. Purushottham Raj and Niessen, F. and Bachmann, F. and Zhang, Y. and Lauridsen, E.},
year = {2026}
}

@article{Sun2017,
author = {Sun, Y.F.},
year = {2017},
month = {05},
pages = {81-92},
title = {Effect of abnormal grain growth on microstructure and mechanical properties of friction stir welded SPCC steel plates},
volume = {694},
journal = {Materials Science and Engineering: A}
}

@InCollection{ParaView,
  author    = {J. Ahrens and B. Geveci and C. Law},
  editor = {C. D. Hansen and C. R. Johnson},
  booktitle = {Visualization Handbook},
  publisher = {Elesvier},
  title     = {{ParaView}: An End-User Tool for Large Data Visualization},
  year      = {2005},
pages = {717-731}
}

@article{zhang_three-dimensional_2018,
	title = {Three-dimensional grain growth in pure iron. {P}art {I}. {S}tatistics on the grain level},
	volume = {156},
	pages = {76--85},
	journal = {Acta Materialia},
	author = {Zhang, J. and Zhang, Y. and Ludwig, W. and Rowenhorst, D. and Voorhees, P.W. and Poulsen, H.F.},
	year = {2018}
}

@article{park_ex_2012,
	title = {Ex Situ Observation of Microstructure Evolution During Abnormal Grain Growth in Aluminum Alloy},
	volume = {43},
	pages = {5218--5223},
	journal = {Metallurgical and Materials Transactions A},
	author = {Park, H.-K. and Kang, H.-G. and Park, C.-S. and Huh, M.-Y. and Hwang, N.-M.},
	year = {2012}
}

@article{suk_development_2006,
	title = {Development of Abnormal Grain Growth in Cold Rolled and Recrystallized {AA} 5182 Sheet},
	volume = {116-117},
	pages = {316--319},
	journal = {Solid State Phenomena},
	author = {Suk, H.G. and Shin, E.J. and Huh, M.Y.},
	year = {2006}
}

@article{ludwig_x-ray_2008,
	title = {X-ray diffraction contrast tomography: a novel technique for three-dimensional grain mapping of polycrystals. {I}. {D}irect beam case},
	volume = {41},
	pages = {302--309},
	journal = {Journal of Applied Crystallography},
	author = {Ludwig, W. and Schmidt, S. and Lauridsen, E.M. and Poulsen, H.F.},
	year = {2008}
}

@article{na_misorientation_2016,
	title = {Misorientation angle analysis near the growth front of abnormally growing grains in 5052 aluminum alloy},
	volume = {115},
	journal = {Acta Materialia},
	author = {Na, T.-W. and Park, H.-K. and Park, C.-S. and Park, J.-T. and Hwang, N.-M.},
    pages = {224-229},
	year = {2016}
}

@article{ludwig_three-dimensional_2009,
	title = {Three-dimensional grain mapping by x-ray diffraction contrast tomography and the use of {F}riedel pairs in diffraction data analysis},
	volume = {80},
	pages = {033905},
	journal = {Review of Scientific Instruments},
	author = {Ludwig, W. and Reischig, P. and King, A. and Herbig, M. and Lauridsen, E.M. and Johnson, G. and Marrow, T.J. and Buffière, J.Y.},
	year = {2009}
}

@article{krill_iii_extreme_2023,
	title = {Extreme abnormal grain growth: connecting mechanisms to microstructural outcomes},
	volume = {53},
	pages = {319--345},
	journal = {Annual Review of Materials Research},
	author = {Krill {III}, C.E. and Holm, E.A. and Dake, J.M. and Cohn, R. and Hol\'{i}kov\'{a}, K. and Andorfer, F.},
	year = {2023}
}

@article{engler_impact_2018,
	title = {Impact of chromium on the microchemistry evolution during solidification and homogenization of the Al-Mg alloy {AA}5052},
	volume = {744},
	pages = {561--573},
	journal = {Journal of Alloys and Compounds},
	author = {Engler, O. and Kuhnke, K. and Westphal, K. and Hasenclever, J.},
	year = {2018}
}

@article{Gao_defectsAA5252_2020,
title = {Formation Mechanism of Strip Defects in Anodizing {AA}5252 Aluminium Alloy Sheets},
volume = {326},
pages = {04003},
journal = {MATEC Web of Conferences},
author = {Gao, C. and Jiang, Z. and Liu, Z. and Zhao, P. and Huang, R. and Wu, Y. and Yu, K.},
year = {2020},
}

@article{Gladman1966,
 author = {Gladman, T.},
 journal = {Proceedings of the Royal Society of London. Series A, Mathematical and Physical Sciences},
 pages = {298--309},
 title = {On the theory of the effect of precipitate particles on grain growth in Metals},
 volume = {294},
 year = {1966}
}

@article{paganin_simultaneous_2002,
	title = {Simultaneous phase and amplitude extraction from a single defocused image of a homogeneous object},
	volume = {206},
	pages = {33--40},
	journal = {Journal of Microscopy},
	author = {Paganin, D. and Mayo, S.C. and Gureyev, T.E. and Miller, P.R. and Wilkins, S.W.},
	year = {2002}
}

@article{park_new_2015,
	title = {New understanding of the role of coincidence site lattice boundaries in abnormal grain growth of aluminium alloy},
	volume = {95},
	pages = {220--228},
	journal = {Philosophical Magazine Letters},
	author = {Park, C.-S. and Park, H.-K. and Shim, H.-S. and Na, T.-W. and Han, C.-H. and Hwang, N.-M.},
	year = {2015},
}

@article{humphreys_unified_1997_2,
	title = {A unified theory of recovery, recrystallization and grain growth, based on the stability and growth of cellular microstructures—{II}. {T}he effect of second-phase particles},
	volume = {45},
	pages = {5031--5039},
	journal = {Acta Materialia},
	author = {Humphreys, F.J.},
	year = {1997},
}

@article{smith_grains_1948,
	title = {Grains, phases and interfaces--an interpretation of microstructure},
	volume = {175},
	series = {Transactions},
	pages = {15--51},
	journal = {American Institute of Mining and Metallurgical Engineers},
	author = {Smith, C.S.},
	year = {1948}
}

@article{read_dislocation_1950,
	title = {Dislocation models of crystal grain boundaries},
	volume = {78},
	pages = {275--289},
	journal = {Physical Review},
	author = {Read, W.T. and Shockley, W.},
	year = {1950}
}

@phdthesis{dake_experimental_2020,
  title        = {Experimental Investigations of Microstructural Coarsening in 3{D} using {X}-Ray Microscopy},
  author       = {Dake, J.M.},
  year         = 2020,
  school       = {Ulm  University},
  type         = {PhD thesis}
}

@article{zoofan_phase-contrast_2006,
    author = {Zoofan, B. and Kim, J.-Y. and Rokhlin, S. I. and Frankel, G. S.},
    title = {Phase-contrast {X}-ray imaging for nondestructive evaluation of materials},
    journal = {Journal of Applied Physics},
    volume = {100},
    pages = {014502},
    year = {2006},
}

@article{manohar_five_1998,
	title = {Five decades of the zener equation},
	volume = {38},
	pages = {913--924},
	journal = {{ISIJ} International},
	author = {Manohar, P.A. and Ferry, M. and Chandra, T.},
	year = {1998},
}

@misc{paleo_nabu_2024,
  author = {Paleo, P. and Lesaint, J. and Payno, H. and Mmirone, A. and Vigano, N. and Nemoz, C.},
  title = {Nabu: {ESRF} tompgraphy processing software},
  year = {2019},
  publisher = {GitHub},
  journal = {GitHub repository},
  howpublished = {https://gitlab.esrf.fr/tomotools/nabu}
}

@misc{payno_tomwer_2023,
  author = {Payno, H. and Paleo, P. and Nemoz, C. and Mirone, A. and Gouttenoire, P.-J. and Lesaint, J. and De Nolf, W. and Vincent, T.},
  title = {Tomwer},
  year = {2016},
  publisher = {GitHub},
  journal = {GitHub repository},
  howpublished = {https://gitlab.esrf.fr/tomotools/tomwer}
}

@article{mackenzie_second_1958,
	title = {Second paper on statistics associated with the random disorientation of Cubes},
	volume = {45},
	pages = {229--240},
	journal = {Biometrika},
	author = {Mackenzie, J.K.},
	year = {1958},
}

@article{reischig_advances_2013,
	title = {Advances in {X}-ray diffraction contrast tomography: flexibility in the setup geometry and application to multiphase materials},
	volume = {46},
	pages = {297--311},
	journal = {Journal of Applied Crystallography},
	author = {Reischig, P. and King, A. and Nervo, L. and Vigan\`{o}, N. and Guilhem, Y. and Palenstijn, W.J. and Batenburg, K.J. and Preuss, M. and Ludwig, W.},
	year = {2013}
}

@misc{dragonfly_2022,
     title = {Dragonfly},
     author = {{Comet Technologies Canada Inc.}},
     year = {2022},
     address = {Montreal, Canada},
note={{V}ersion: 2022.2}
}

@book{humphreys_recrystallization_2017,
Author = {Humphreys, J. and Rohrer, G.S. and Rollett, A.},
Title = {Recrystallization and Related Annealing Phenomena},
edition={3rd},
publisher = {Elsevier},
year = {2017}
}

@article{lu_dynamics_2020,
	title = {Dynamics of particle-assisted abnormal grain growth revealed through integrated three-dimensional microanalysis},
	volume = {195},
	pages = {1--12},
	journal = {Acta Materialia},
	author = {Lu, N. and Kang, J. and Senabulya, N. and Keinan, R. and Gueninchault, N. and Shahani, A.J.},
	year = {2020}
}

@article{dennis_abnormal_2009,
	title = {Abnormal grain growth in {A}l–3.5{C}u},
	volume = {57},
	pages = {4539--4547},
	journal = {Acta Materialia},
	author = {Dennis, J. and Bate, P.S. and Humphreys, F.J.},
	year = {2009}
}

@article{dake_2016,
author = {Dake, J.M. and Oddershede, J. and S{\o}rensen, H.O. and Werz, T. and Shatto, J.C. and Uesugi, K. and Schmidt, S. and Krill {III}, C.E},
title = {Direct observation of grain rotations during coarsening of a semisolid {A}l–{C}u alloy},
journal = {Proceedings of the National Academy of Sciences},
volume = {113},
pages = {E5998-E6006},
year = {2016}
}

@article{lienert_high-energy_2011,
	title = {High-energy diffraction microscopy at the advanced photon source},
	volume = {63},
	pages = {70--77},
	journal = {JOM: The Journal of the Minerals, Metals \& Materials Society},
	author = {Lienert, U. and Li, S.F. and Hefferan, C.M. and Lind, J. and Suter, R.M. and Bernier, J.V. and Barton, N.R. and Brandes, M.C. and Mills, M.J. and Miller, M.P. and Jakobsen, B. and Pantleon, W.},
	year = {2011}
}

@article{rollet_texture_AGG_2005,
  author = {A. Rollett},
  title = {Abnormal grain growth and texture development},
  year = {2005},
  journal = {Materials Science Forum},
  volume = {495-497},
  pages = {1171-1176},
}

@article{Kim_Al_texture_AGG_2007,
title = {Effect of primary recrystallization texture on abnormal grain growth in an aluminum alloy},
journal = {Scripta Materialia},
volume = {57},
pages = {325-327},
year = {2007},
author = {H.-C. Kim and C.-G. Kang and M.-Y. Huh and O. Engler},
}

@article{Wilhelm2026,
  title={A computational framework for tracking grain boundaries in 3{D} image data: Quantifying boundary curvatures and velocities in polycrystalline materials},
 journal = {Computational Material Science},
 author={Wilhelm, T. and Furat, O. and Dake, J.M. and Krill {III}, C.E. and Schmidt, V.},
 volume = {267},
 pages={114541},
 year={2026}
}

@article{Furnace1999,
    author = {Margulies, L. and Kramer, M. J. and McCallum, R. W. and Kycia, S. and Haeffner, D. R. and Lang, J. C. and Goldman, A. I.},
    title = {New high temperature furnace for structure refinement by powder diffraction in controlled atmospheres using synchrotron radiation},
    journal = {Review of Scientific Instruments},
    volume = {70},
    pages = {3554-3561},
    year = {1999},
}

@article{gladman_abnormal_1992,
	title = {Abnormal Grain Growth during the Heat Treatment of Steel},
	volume = {94-96},
	pages = {113--128},
	journal = {Materials Science Forum},
	author = {Gladman, T.},
	year = {1992},
}

@article{DeLellis_AGGalumina_2026,
title = {3{D} analysis of abnormal grain growth in calcia doped alumina in the presence of large pores},
journal = {Journal of the European Ceramic Society},
volume = {46},
pages = {118173},
year = {2026},
issn = {0955-2219},
doi = {https://doi.org/10.1016/j.jeurceramsoc.2026.118173},
author = {D. P. DeLellis and Y. Wang and V. Muralikrishnan and J. B. Harley and M. R. Tonks and A. R. Krause} 
}

@article{holm_particle-assisted_2015,
	title = {Particle-assisted abnormal grain growth},
	volume = {89},
	issn = {1757-899X},
	url = {https://doi.org/10.1088/1757-899x/89/1/012005},
	doi = {10.1088/1757-899X/89/1/012005},
	pages = {012005},
	journal = {{IOP} Conference Series: Materials Science and Engineering},
	shortjournal = {{IOP} Conf. Ser.: Mater. Sci. Eng.},
	publisher = {{IOP} Publishing},
	author = {Holm, E. A. and Hoffmann, T. D. and Rollett, A. D. and Roberts, C. G.},
	year = {2015},
	langid = {english},
}

@article{Fang2025,
author = "Fang, H. and Ludwig, W.",
title = "{Forward-model-based grain reconstruction to improve the tolerance of diffraction contrast tomography for increased sample deformation}",
journal = "Journal of Applied Crystallography",
year = "2025",
volume = "58",
pages = "796--812",
}

@incollection{AGG_inconel718,
  title={Characterization of microstructures containing abnormal grain growth zones in alloy 718},
  author={J. Cormier and
  O. Yousfi and
  P. Villechaise and
  Y. Dahan and
  B. Flageolet},
  booktitle={Proceedings of the Conference on Superalloys},
  year={2010},
  pages={594-606},
  url={https://api.semanticscholar.org/CorpusID:137237238}
}

@article{Herring1951a,
  title={Some theorems on the free energies of crystal surfaces},
  author={C. Herring},
  journal={Physical Review},
  volume={82},
  pages={87},
  year={1951},
  publisher={APS}
}

@incollection{Herring1951b,
  author    = {C. Herring},
  title     = {Surface tension as a motivation for sintering},
  booktitle = {The Physics of Powder Metallurgy},
  editor    = {Kingston, W. E.},
  pages     = {143--179},
  year      = {1951},
  publisher = {McGraw-Hill},
  address   = {New York}
}

@article{labDCTHolzner_2016,
author = {Holzner, C. and Lavery, L. and Bale, H. and Merkle, A. and McDonald, S. and Withers, P. and Zhang, Y. and Jensen, D. and Kimura, M. and Lyckegaard, A. and Reischig, P. and Lauridsen, E.},
year = {2016},
pages = {34-43},
title = {Diffraction contrast tomography in the laboratory – applications and future directions},
volume = {24},
journal = {Microscopy Today}
}

@article{ouhiba_6016_aluminium,
  title = {Formation of coarse recrystallized grains in 6016 aluminum alloy during holding after hot deformation},
  author = {Ouhiba, S. and Nicolaÿ, A. and Boissonnet, L. and Bernacki, M. and Bozzolo, N.},
  journal = {Metallurgical and Materials Transactions A},
  publisher = {Springer Verlag/ASM International},
  volume = {53},
  pages = {2402-2425},
  year = {2022}
}

\clearpage

\begin{titlepage}
\centering

\vspace*{3cm}

{\large\bfseries Supplementary material for:\par}

\vspace{1cm}

{\LARGE
Tracking the Emergence and Persistence of Abnormal Grain Growth in an Aluminum Alloy using Diffraction-Contrast Tomography
\par}

\vspace{1.5cm}

{\large
Helmuth-André Schulz-Harder$^1$, Karolína Gutbrod$^1$, Madlen Atzen$^1$,
Markus Ziehmer$^1$, Thomas Wilhelm$^2$, Orkun Furat$^{2,3}$, 
Yubin Zhang$^4$, Jette Oddershede$^5$,
Haixing Fang$^6$, Pierre-Olivier Autran$^6$, Wolfgang Ludwig$^{6,7}$,
Marc Bernacki$^8$, Volker Schmidt$^2$, Carl E. Krill III$^1$, Jules M. Dake$^1$
\par}

\vspace{1cm}

{\normalsize
$^1$Institute of Functional Nanosystems, Ulm University, 89081~Ulm, Germany\\
$^2$Institute of Stochastics, Ulm University, 89069~Ulm, Germany\\
$^3$SDU Applied AI and Data Science Unit, University of Southern Denmark,
5230~Odense, Denmark\\
$^4$Department of Civil and Mechanical Engineering, Technical University of Denmark, DK-2800 Kgs. Lyngby, Denmark\\
$^5$Xnovo Technology ApS, Galoche Allé 15, 1st floor, 4600~Køge, Denmark\\
$^6$European Synchrotron Radiation Facility, 71 Avenue des Martyrs,
3800~Grenoble, France\\
$^7$Université de Lyon, INSA Lyon, CNRS MATEIS,
69621~Villeurbanne, France\\
$^8$Mines Paris, PSL University, Centre for Material Forming (CEMEF),
UMR CNRS, 06904~Sophia Antipolis, France
\par}

\vfill

\end{titlepage}

\setcounter{figure}{0}
\renewcommand{\thefigure}{S\arabic{figure}}

\clearpage
\section*{Annealing and Measurement Protocol}\label{app:protocol}
\begin{figure}[H]
    \centering
    \includegraphics[width=0.85\linewidth]{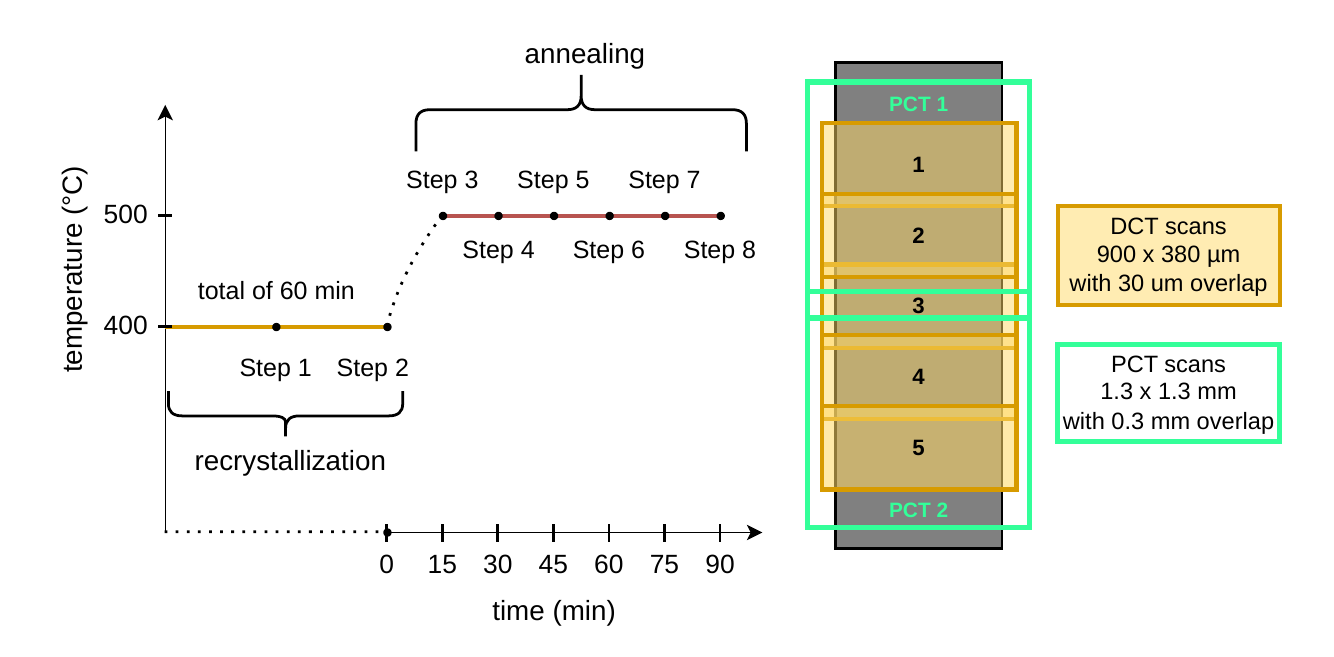}
    \caption{Complete annealing and measurement protocol for the AA5252 sample. After full recrystallization for $\SI{60}{\minute}$ at $\SI{400}{\celsius}$, the sample was annealed in 15-minute steps at $\SI{500}{\celsius}$ for a total of $\SI{90}{\minute}$. To cover a larger sample volume, the sample was characterized by five overlapping DCT scans and two overlapping PCT scans at room temperature after each annealing step.}
    \label{fig:append:protocol}
\end{figure}

\clearpage
\section*{Raw Diffraction Images}\label{app:raw_diff_ims}
\begin{figure}[h!]
     \centering
     \includegraphics[width=0.75\textwidth]{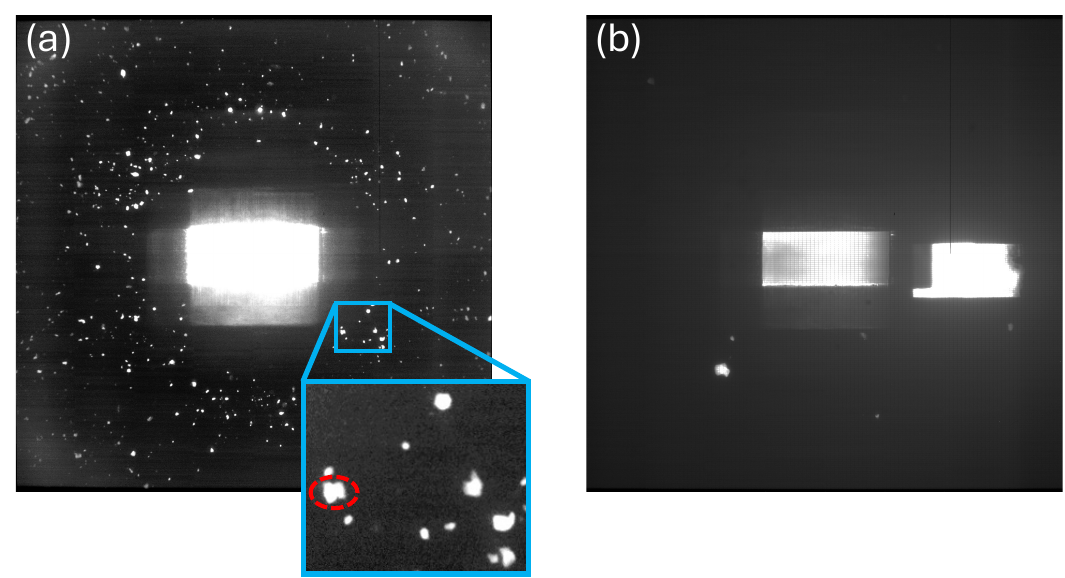}
     \caption{(a) Diffraction image recorded at a single rotation angle during the DCT scans, showing diffraction spots from thousands of equally sized matrix grains during the initial stages of annealing. The diffraction spots frequently overlap, as illustrated by the red-circled region in the lower-right inset. (b) Diffraction image recorded at a later time step, showing a few diffraction spots from matrix grains alongside a prominent diffraction spot originating from an abnormal grain. The latter is partially truncated by the beam stop at the center of the image.}
     \label{fig:append:diffspots}
\end{figure}

\clearpage
\section*{Influence of the Bin Width on the Particle Clustering Effect}
\begin{figure}[H]
    \centering
    \captionsetup{width=1\linewidth}
   \includegraphics[width=1\linewidth]{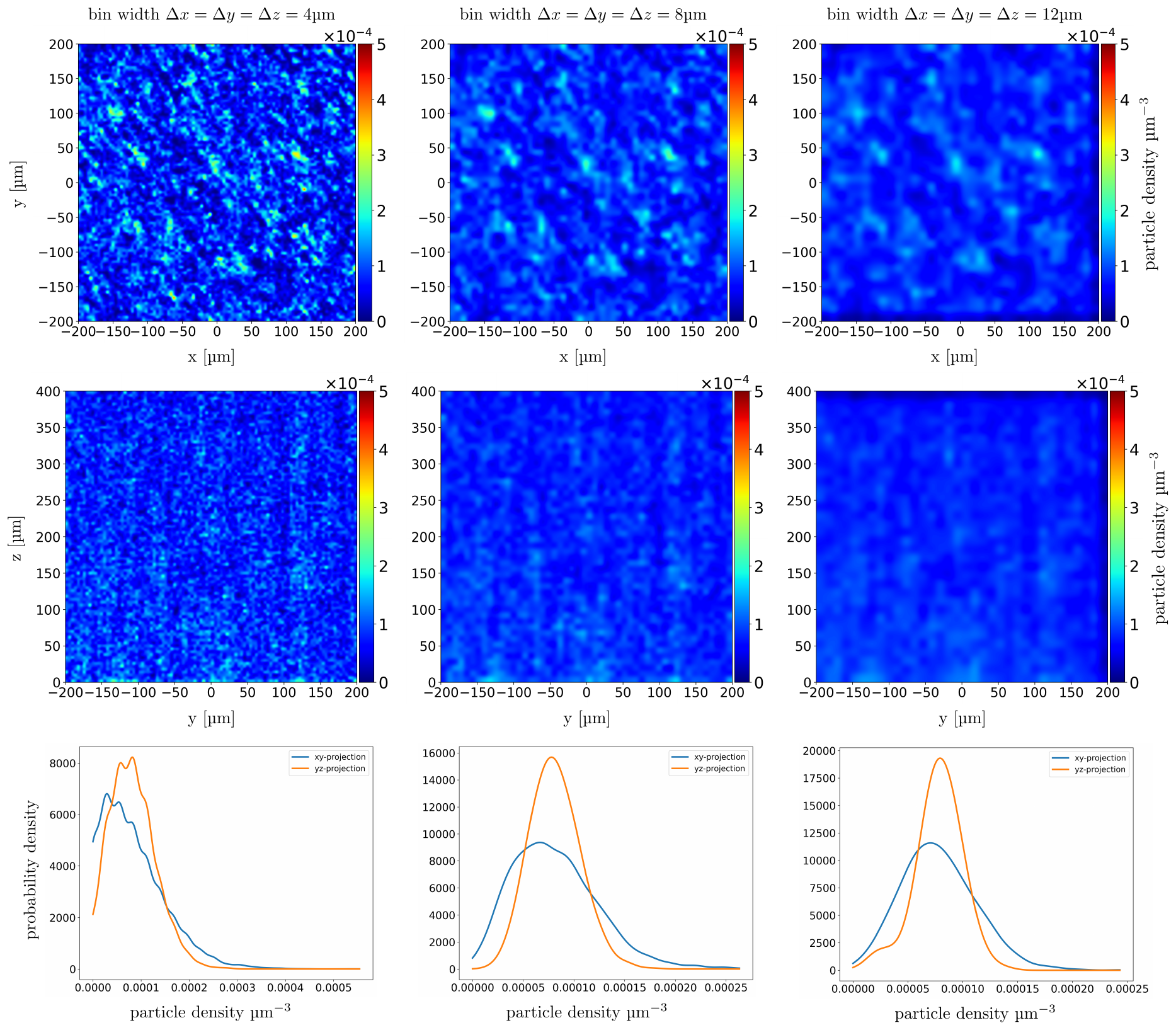}
    \caption{Analysis of the spatial frequency of pinning particles with respect to the $(x,y,z)$ coordinates of their center positions in the cylindrical sample at $\SI{15}{\minute}$. In the first two rows, the density of particle center positions, projected onto (top row) the $x$-$y$ plane and (middle row) the $y$-$z$ plane, is shown for three different bin widths ($\SI{4}{\micro\meter}$, $\SI{8}{\micro\meter}$ and $\SI{12}{\micro\meter}$). In the bottom row, kernel density estimates illustrate the distribution of particle number densities in the figures plotted in the top and middle rows.} 
    \label{fig:append:particles_2}
\end{figure}

\end{document}